\documentclass[showpacs,amssymb,superscriptaddress,notitlepage,nofootinbib,prd,longbibliography]{revtex4-2}
\usepackage{graphicx}
\usepackage{amsmath}
\usepackage{amsfonts}
\usepackage{bm}
\usepackage[colorlinks=true]{hyperref}
\usepackage[all]{hypcap}
\usepackage[utf8]{inputenc}
\usepackage{url}

\usepackage{float}

\usepackage{xcolor}
\usepackage{multirow}
\usepackage{ebgaramond}
 \usepackage{etoolbox}

\hypersetup{
    colorlinks,
    citecolor=[rgb]{0.16,0.53,0.80},
    linkcolor=[rgb]{0.16,0.03,0.50},
}

\input colordvi
\usepackage{tikz}
\usepackage{graphicx,color}

\def\be{\begin{equation}}
\def\ee{\end{equation}}
\def\bea{\begin{eqnarray}}
\def\eea{\end{eqnarray}}
\def\ptl{\partial}

\begin{document}

\thispagestyle{empty}

\centerline{\Large{ \bf Rotational Curves of the Milky Way Galaxy
and Andromeda Galaxy  }}

\medskip

\centerline{\Large{ \bf  in Light of Vacuum Polarization around
Eicheon}} \bigskip \centerline{S.L. Cherkas $^a$, V.L.
Kalashnikov$^b$}
\medskip
\centerline{$^a$Institute for Nuclear Problems, Belarus State
University} \centerline{Minsk 220006, Belarus} \centerline{Email:
cherkas@inp.bsu.by}
\medskip
\centerline{$^b$ Department of Physics,}\centerline{  Norwegian
University of Science and
Technology,}\centerline{H{\o}gskoleringen 5, Realfagbygget, 7491,
Trondheim, Norway} \centerline{Email:
vladimir.kalashnikov@ntnu.no}
\medskip
\centerline{\small Submitted: September 21, 2023}

\smallskip

{\fontsize{10}{11}\selectfont Eicheon properties  are discussed.
It is shown that the eicheon surface allows setting a boundary
condition for the vacuum polarization and obtaining a solution
describing the dark matter tail in the Milky Way Galaxy. That is,
the dark matter in the Milky Way Galaxy is explained as the F-type
of vacuum polarization, which could be treated as dark radiation.
The model presented is spherically symmetric, but a surface
density of a baryonic galaxy disk is taken into account
approximately by smearing the disk over a sphere. This allows the
reproduction of the large distance shape of the Milky Way Galaxy
rotational curve. Andromeda Galaxy's rotational curve is also
discussed.}

\section{Introduction}

Observation of the stellar orbits around the center of the Milky
Way Galaxy~\cite{Gillessen2009,Nampalliwar2021}, detecting the
gravitational waves from the black hole/black hole and black
hole/neutron star coalescence (e.g., see the
catalog~\cite{abbott2021gwtc} for an overview), radio-astronomy
observation of the ``black hole shadows'' in the centers of
galaxies~\cite{akiyama2019first,johnson2018scattering}  are widely
considered as the direct evidence of an extremely compact
astrophysical object (ECO) existence with a radius of an order of
the Schwarzschild one. The~observable properties of such an object
are well-described by an exact Schwarzschild (or, more precisely,
Kerr) solution of the general relativity (GR)
equations~\cite{lan,chandra}. A~principal question is whether the
Schwarzschild solution interprets reality quite adequately.
Indeed, there are a lot of theoretical attempts to describe ECO
whose properties approach those of an ordinary GR black hole
sufficiently far from the event horizon (so-called horizonless
``exotic compact objects'' \cite{cardoso2019testing}). Some of
them are based on the modified theories of gravity\footnote{For
the comprehensive reviews,
see~\cite{berti2015testing,yagi2016black}, and~one may add an
additional alternative approach based on the relativistic theory
of gravity with a massive
graviton~\cite{logunov1997possibility}.}. Recently, ECOs without a
horizon have been discussed intensively
(e.g.,~\cite{PhysRevD.77.044032,chapline,carballorubio2022connection,carballorubio2023singularityfree}).
A zoo of exotic ECO, such as bosonic
stars~\cite{schunck2003general},
gravastars~\cite{ray2020gravastar}, and~other exotic
stars~\cite{urbano2019gravitational,singh2020compact}, was
proposed and theoretically explored. Also, the~approaches based on
constructing the nonsingular black-hole metrics in the spacetimes
of different dimensions were proposed (e.g.,
see~\cite{hayward2006formation,frolov2016notes}).

The question about the nature of ECO is also related to the need
for dark matter (DM) to explain the galactic rotational
curves~\cite{Sofue,baes2003observational,Dai2022}. In~particular,
the~first observation of the DM density around the stellar-mass
ECO appears~\cite{chan2023indirect}. It was conjectured that the
primordial black holes could be considered the candidates to
DM~\cite{carr2020primordial}. Besides, there is a plethora of DM
candidates~\cite{bertone2018new}. However, could we advance
without extraordinary physics but only by taking a vacuum
polarization into account correctly~\cite{polar2022}? Conventional
answer is ``No'' in the frame of the renormalization technique of
quantum field theory on a curved
background~\cite{Birrell82,Brunetti2009}. Still, this approach
demands covariance of the mean value of the energy-momentum tensor
over the vacuum state~\cite{Birrell82}. This demand has no solid
foundation because it is known that there is no vacuum state
invariant relative to the general transformation of coordinates.
On the contrary, an~argument was put forward that the preferred
conformally-unimodular metric (CUM) could describe a vacuum
polarization and resolve the DM
problem~\cite{polar2022,Cherkas2022}. In~this metric, a~black hole
as an object having a horizon is absent~\cite{Eicheons}. Its
disappearance results from the coordinate transformation relating
the Schwarzschild-type metric to CUM, which selects some shell
over the horizon and draws it into a node. As~a result, a~point
mass without a horizon arises in a CUM. It is an idealized
picture. In~reality, one must know the equation of the state of a
substance forming such ECO (named ``eicheon'' \cite{Eicheons}).

Here, aiming at understanding the eicheon nature, we will use an
approximation of the constant energy density and a trial
``equation of state'' relating the maximal pressure and the energy
density.  In~our approach, we construct eicheon, and,
after~determining its properties, describe an eicheon surrounded
by ``dark radiation'' to explain the rotational curves of the
Milky Way  and Andromeda  (M31) Galaxies. ``Dark radiation'' is
one of two kinds of vacuum polarization considered
in~\cite{polar2022}, namely the polarization of F-type. Finally,
to~be closer to observations, we introduce a baryonic matter into
the model by smearing the galactic disks of the Milky Way Galaxy
and~M31.

\section{What Is ``Eicheon''?}

\begin{figure}[H]
  \includegraphics[width=11cm]{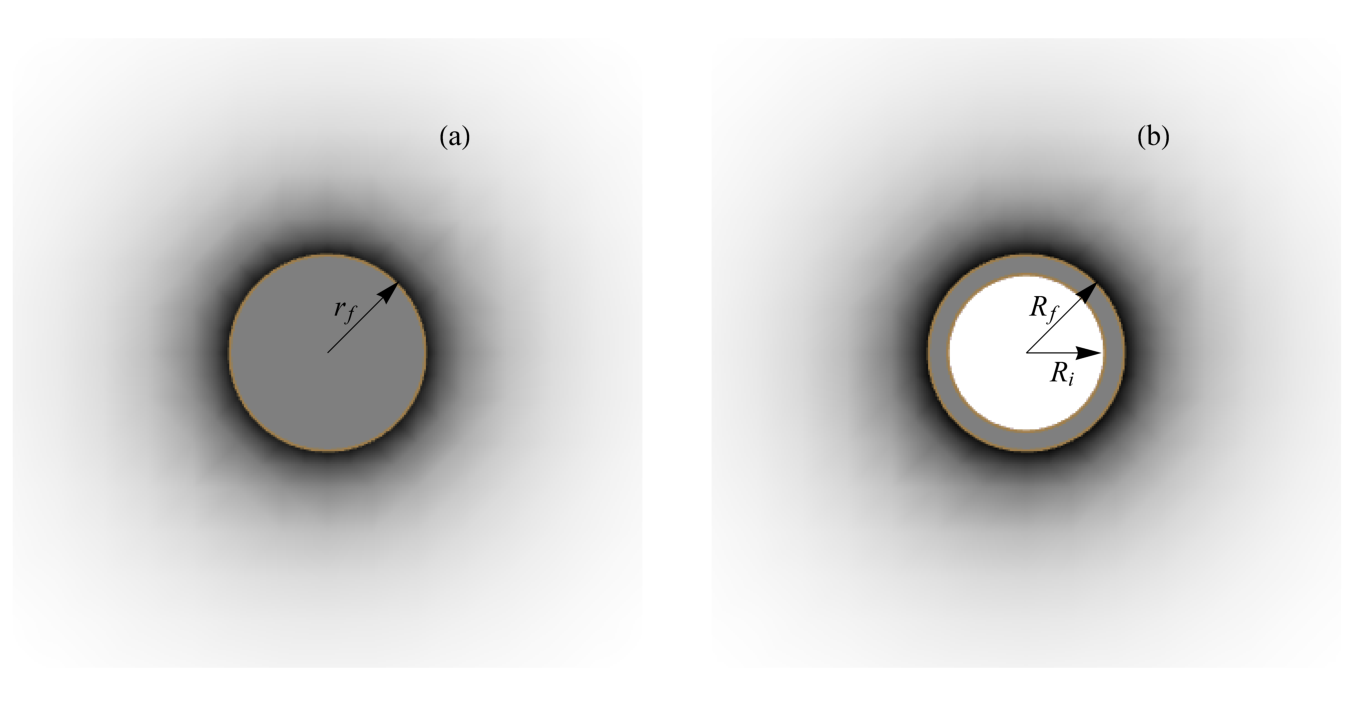}
  \caption{(\textbf{a}) Nonsingular eicheon surrounded by dark radiation in CUM
(\ref{eq42}) has a nonsingular core. (\textbf{b}) In the
Schwarzschild type metric (\ref{ab}), this core looks like a
hollow
  sphere. Vacuum polarization around an eicheon is shown as
  the gradient of a density.}
\label{fig:piceich}
\end{figure}

Eicheon is a horizon-free object which appears instead of a black
hole in CUM. As~an idealized structure, eicheon represents
 a solution of a gravitational field of a point mass
in CUM. In~the metric of a Schwarzschild type, it looks like a
massive shell situated over the Schwarzschild radius. In~the real
world, where there is no infinite density and pressure,
the~eicheon could be modeled in the Schwarzschild-type metric by a
layer of finite width over the horizon, as~it is shown in
Figure~\ref{fig:piceich}. In~CUM, it looks like a solid
ball~\cite{Eicheons,v2}. A~constant density model is convenient
for understanding the main features of the~eicheon.

CUM for a spherically symmetric space-time is written as
\begin{equation}
  ds^2 = a^2(d\eta ^2 - \tilde {\gamma }_{ij} dx^idx^j) =
e^{2\alpha }\left( {d\eta ^2 - e^{ - 2\lambda }(d{\bm x})^2 -
(e^{4\lambda } - e^{ - 2\lambda })({\bm x}d{\bm x})^2 / r^2}
\right),
\label{eq31a}
\ee
where  $r = \vert \bm x \vert $, $a = \exp \alpha $, and~$\lambda
$ are the functions of $\eta , r$.  The~matrix  $\tilde
\gamma_{ij}$ with the unit determinant is expressed through
$\lambda(\eta,r)$. The~interval (\ref{eq31a}) could be also
rewritten in the spherical coordinates:
\bea
x=r\sin \theta\cos\phi, ~~y=r\sin \theta\sin\phi,~~ z=r\cos\theta
\eea
resulting in
\begin{equation}
ds^2 = e^{2\alpha }\left( {d\eta ^2 - dr^2e^{4\lambda } - e^{ -
2\lambda }r^2\left( {d\theta ^2 + \sin^2\theta d\phi ^2} \right)}
\right){\kern 1pt} .
\label{eq42}
\end{equation}

 However, let us discuss eicheon properties in
the  Schwarzschild-type metric, which is more convenient for a
reader
\be
ds^2=B(R)dt^2-A(R)dR^2-R^2d\Omega.
\label{ab}
\ee

In this metric, the~Volkov-Tolman-Oppenheimer (TOV) equation for a
layer $R\in\{R_i,R_f\}$  reads as:
\begin{equation}
p^\prime(R)=-\frac{3}{4\pi M_p^2R^2} \mathcal M(R) \rho (R)
\left(1+\frac{4 \pi R^3 p(R)}{\mathcal M(R)}\right)
\left(1+\frac{p(R)}{\rho (R)}\right)\left(1-\frac{3 \mathcal
M(R)}{2\pi M_p^2 R}\right)^{-1},
\label{op}
\end{equation}
where the function
\be\mathcal M(R)=4\pi \int_{R_i}^{R}
\rho(R^\prime) R^{\prime 2}dR^\prime
\label{mmr}
\ee
and the reduced Planck mass $M_p=\sqrt{\frac{3}{4\pi G
}}=1.065\times 10^{-8}~kg$. We will model a layer of constant
density $\rho$ so that $\mathcal M(R)$ is expressed as
\be
\mathcal M(R)=\frac{4\pi}{3}\rho\left(R^3-R_i^3\right).
\label{mr}
\ee

 {It} 
 is convenient to measure distances in units of the
Schwarzschild radius $r_g=\frac{3\mathbb{M}}{2\pi M_p^2}$, while
energy density and pressure in the units of $M_p^2r_g^{-2}$.
In~these units, it follows from (\ref{mr}) and the definition of
the eicheon mass  $ \mathbb{M}=\mathcal M(R_f)$ that
\be
\rho=\frac{1}{2(R_f^3-R_i^3)}.
\label{rinun}
\ee

 {The} TOV Equation~(\ref{op}) is reduced to
\be
p'=\frac{(p+\rho ) \left(3p R^3+\rho
\left(R^3-R_i^3\right)\right)}{R \left(2 \rho
\left(R^3-R_i^3\right)-R\right)}
\label{tov0}
\ee
and has to be solved with the boundary condition
$p(R_f)=p\left(\sqrt[3]{R_i^3+\frac{1}{2\rho}}\right)=0$, where
the second equality follows from (\ref{rinun}). Let us simplify
the problem further and assume that $R_i=1$ in the Schwarzschild
radius units. Even in this case, there is no analytical solution
of the Equation~(\ref{tov0}), but~the most interesting quantity is
a maximal pressure $p_{max}=p(1)$, which turns out to be
approximated by the expression
\be
p_{max}\approx\frac{\sqrt{\rho
}}{\sqrt{6}}-\frac{1}{3}+\frac{11}{36 \sqrt{6} \sqrt{\rho
}}-\frac{35}{864 \sqrt{6} \rho ^{3/2}}
\label{approx}
\ee
as is shown in Figure~\ref{fig:fig0}.

\begin{figure}[H]
\includegraphics[width=8.5cm]{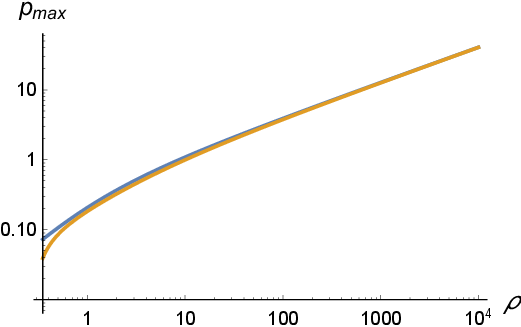}
    \caption{Pressure $p_{max}$ in the center of eicheon (see Figure~\ref{fig:piceich}a) in CUM, coinciding with the pressure $p(R_i)$ in the metric (\ref{ab}) (see Figure~\ref{fig:piceich}b).   Blue and brown curves correspond to the numerical integration of the Equation~(\ref{tov0}) and approximation (\ref{approx}), respectively.}
    \label{fig:fig0}
\end{figure}

If one supplements Equation~(\ref{approx}) by the ``equation of
state'', which connects the maximal pressure with the density,
then it is possible to determine the pressure and density.
For~instance, the~``equation of state''
  corresponding to a degenerate relativistic fermion gas
  \be
p_{max}=\rho/3
\label{radeq}
  \ee
  gives no solution because of Equations~(\ref{approx}) and
  (\ref{radeq}) are~incompatible.

The equation of state of the nonrelativistic degenerate Fermi gas
is written in physical units as~\cite{Landau1980}

\be
\tilde p_{max} =
\frac{1}{5}\left(\frac{3\pi^2}{m_N^4}\right)^{2/3}\tilde
\rho^{5/3},
\label{elgas}
\ee
where $m_N$ is a particle mass, and~the tilde denotes that the
quantity is expressed in the  physical units. When $\rho$ is
large, one could use only the first term in
Equation~(\ref{approx}), and its equating to the pressure from
(\ref{elgas}) gives the following expression
  \be
\frac{{M_p r_g^{-1}} \sqrt{\tilde\rho }}{\sqrt{6} }=\frac{3^{2/3}
\pi ^{4/3}}{5}
 \left(\frac{1}{m_N^4}\right)^{2/3} \tilde\rho
^{5/3},
  \ee
allowing us to find the physical density
\be
\tilde \rho=\frac{\left(\frac{5}{3}\right)^{6/7} 2^{3/7}
{m_N}^{16/7} {M_p}^{18/7}}{3 \pi ^{2/7} \mathbb{M}^{6/7}},
\label{physden}
\ee
which decreases with  an increase of mass $\mathbb{M}$ of the
eicheon. Dimensionless density is found by dividing
(\ref{physden}) by $M_p^2r_g^{-2}$ and reads
\be
\rho=\frac{\sqrt[7]{3}\, 5^{6/7} \mathbb{M}^{8/7}m_N^{16/7}}{
2^{11/7} \pi ^{16/7} M_p^{24/7}}.
\ee

 {It} grows with the increase of $\mathbb{M}$, so that
approximation $p_{max}\approx \sqrt{\rho/6}$ becomes justified at
some mass according to (\ref{approx}). Respectively, the~width of
the eicheon shell decreases: $\Delta
R=\sqrt[3]{1+\frac{1}{2\rho}}-1\approx \frac{1}{6\rho}$ and
becomes very thin at large $\mathbb{M}$. Certainly, we imply the
relative width in units of $r_g$. For~instance, if~one takes the
eicheon mass equal to the Sun mass
$\mathbb{M}=M_\odot=1.989\times10^{30}$~kg and $m_N$ equals the
neutron mass, then the dimensional density $\tilde \rho=2.4\times
10^{19}$~kg/m$^3$, while the dimensionless $\rho$ equals $0.66$.
This eicheon has a rather thick skin $\Delta R\approx 0.33$ and,
in the principle can be distinguished from a conventional black
hole. One more example is the eicheon of a large mass $40\times
10^9 M_\odot$. In~this case, the~physical density is much lower
and we could consider the ``equation of state'' for a cold
hydrogen plasma, where the pressure is created by a degenerate
electron gas, and~the dimensional density satisfies
\be
\frac{M_p r_g^{-1} \sqrt{\tilde \rho }}{\sqrt{6} }=\frac{3^{2/3}
\pi ^{4/3}}{5}
 \left(\frac{1}{m_e^4}\right)^{2/3}
\left(\frac{\tilde \rho \,m_e }{m_N}\right)^{5/3},
\label{f1}
\ee
where $m_e$ is electron mass and ${\tilde \rho \,m_e }/{m_N}$ is
electron density in a fully ionized  hydrogen plasma. Thus
\be
\tilde \rho=\frac{\left(\frac{5}{3}\right)^{6/7} 2^{3/7}
{m_e}^{6/7} {m_N}^{10/7} {M_p}^{18/7}}{3 \pi
^{2/7}\mathbb{M}^{6/7}}.
\label{fizfiz}
\ee

 {The} dimensionless density is given by
\be
 \rho=\frac{\sqrt[7]{3}\, 5^{6/7} \mathbb{M}^{8/7} {m_e}^{6/7}
{m_N}^{10/7}}{2^{11/7} \pi ^{16/7} {M_p}^{24/7}}.
\label{f2}
\ee

 {Numerically,} these values are $\tilde \rho=3.1\times
10^7$~kg/m$^3$, $\rho=1.4\times 10^{9}$. The~eicheon skin is very
thin $\Delta R\sim \frac{1}{6\rho}\sim 10^{-10}$. Such eicheon is
indistinguishable from a conventional black hole. At~the same
time, it is rather ``mellow'' by virtue of (\ref{fizfiz}).
Certainly, there is no paradox here because $\Delta R$ is measured
in the units of $r_g$, which is large in the case considered.
Finally, we can estimate eicheon in the center of the Milky Way
Galaxy using the Formulas (\ref{f1})--(\ref{f2}). For
$\mathbb{M}=4.154\times 10^6 M_\odot$, they give $\rho\approx
3.8\times 10^4$, $\Delta R\sim 10^{-6}$ and $\tilde\rho\approx
8.2\times 10^{10}$~kg/m$^3$ that is greater than the white dwarf
mean density $\tilde\rho\approx 4\times
10^8$~kg/m$^3$~\cite{weidemann1968white}. The~eicheons of any mass
exist because the inner $R_i$ and outer $R_f$ radii (see
Figure~\ref{fig:piceich}b ) exceed the Schwarzschild radius,
and~Buchdahl's bound~\cite{Buchdahl} $\mathbb{M}<4R/9G$ is not
eligible.

To consider eicheon in CUM (\ref{eq42}), one could set $t=\eta$,
$R=R(r)$ and compare the metrics  (\ref{eq42}) and (\ref{ab}) to
obtain:
\begin{equation}
B(R)=e^{2\alpha },~~~~~~~~~~~~~~~~~~~\label{e1}
\end{equation}
\begin{equation}
R^2=r^2e^{-2\lambda+2\alpha},~~~~~~~~~~~~~~~~\label{e2}
\end{equation}
\begin{equation}
 A(R)\left(\frac{dR}{dr}\right)^2=e^{4\lambda+2\alpha}\label{e3}.
\end{equation}

Using (\ref{e1}) and (\ref{e2}) in (\ref{e3}) to exclude $\lambda$
and $\alpha$ yields
\begin{equation}
\label{eqr}
\frac{dr}{dR}=\frac{R^2}{r^2}\frac{A^{1/2}}{B^{3/2}}.
\end{equation}

In the region filled by matter, $A(R)$ and $B(R)$ obey
~\cite{wein}\vspace{-6pt}
\bea
\label{eqA}
\frac{d}{dR}\left(\frac{R}{A}\right)=1-{6}\rho R^2,\nonumber\\
\label{eqB}
\frac{1}{B}\frac{dB}{dR}=-\frac{2}{p+\rho}\frac{dp}{dR}.
\eea

 {For} the model of a constant density $\rho(R)=const$,
the~Equation~(\ref{eqB}) can be integrated explicitly
\begin{equation}
A=\frac{R}{R-1-2 \rho \left(R^3-R_f^3\right)}.
\end{equation}
\begin{equation}
B=\left(1-\frac{1}{R_f}\right)\frac{\rho^2}{(p(R)+\rho)^2}\approx
1-\frac{1}{R_f},
\label{exB}
\end{equation}
 where the pressure is neglected compared to the energy density in the last equality of (\ref{exB}).
 According to  (\ref{eqr}), the~eicheon radius is
\begin{equation}
r_f=\sqrt[3]{3\int_1^{R_f}\frac{A^{1/2}}{B^{3/2}}R^2dR} \approx
\frac{\sqrt{3} \sqrt[3]{11} \,\rho^{1/6}}{2^{5/6}}+\frac{43}{
2^{5/6}\, {3^{5/2}} \,11^{2/3}\rho ^{5/6}},
\label{rff0}
\end{equation}
where a small ``thickness'' of the eicheon  surface $R_f-1$ is
assumed, and~$R_f$ is expressed as
$R_f=\sqrt[3]{1+\frac{1}{2\rho}}$ . For~a supermassive eicheon,
using (\ref{f2}) and first term of (\ref{rff0}) results
\be
r_f\approx \frac{3^{11/21} \sqrt[7]{5} \sqrt[3]{11} }{ 2^{23/21}
\pi ^{8/21}}\sqrt[42]{\frac{\mathbb{M}^8 m_e^6
m_N^{10}}{M_p^{24}}},
\ee
i.e., in~CUM, the~eicheon radius in the units of $r_g$ increases
when the eicheon mass $\mathbb{M}$ rises.

\section{Vacuum Polarization around of~Eicheon}

Considering the vacuum polarization for an arbitrary curved
space-time background is a highly complex problem. Instead, one
could consider the scalar perturbations of CUM:
\be
  ds^2=(
1+\Phi(\eta,\bm
x))^2\left(d\eta^2-\left(\left(1+\frac{1}{3}\sum_{m=1}^3
\ptl_m^2F(\eta,\bm x)\right)\delta_{ij}-\ptl_i\ptl_jF(\eta,\bm
x)\right)dx^idx^j\right)
\label{int1}
\ee
and calculate a spatially nonuniform energy density and pressure
arising due to vacuum polarization in the eikonal
approximation~\cite{polar2022}.

As was shown~\cite{polar2022}, the~energy density and pressure of
vacuum polarization corresponding to the F-type of metric
perturbations (\ref{int1}) have the radiation equation of state
$\delta p_F=\frac{1}{3}\delta \rho_F $. ``Dark radiation''
 does not consist of real particles\footnote{One could imagine, that ``dark radiation'' consists
of virtual particles of some kind.  We have
considered~\cite{polar2022} only scalar field, minimally coupled
with gravity.}, nor interacts with some substance, but~could be a
source in the equations for gravity. That gives a possibility to
use a hypothetical ``dark radiation'' in some heuristic nonlinear
models, such as the TOV equation. For~a radiation substance alone,
a singular solution of the TOV equation exists that is devoid of
physical meaning~\cite{wein}. However, the~situation changes
cardinally in CUM in the presence of the nonsingular eicheon. This
gives a possibility to set a boundary condition for a radiation
fluid at $r=0$ and obtain a nonsingular solution, including the
dark radiation. In~the Schwarzschild type metric (\ref{ab}),
the~boundary condition is set at the radial coordinate of an inner
shell $R=R_i$, which corresponds to the point $r=0$ in CUM (see
Figure~\ref{fig:piceich}).

The system of equations (see Appendix \ref{dertov}) in the metric
(\ref{ab}), implies three substances: the eicheon of the constant
density $\rho_1$, the~dark radiation density $\rho_2$, and~the
density $\rho_3$ of  baryonic matter  of the galactic disk and
bulge:
\be
 \left\{ {{\begin{array}{*{20}c}
 \left\{ {{\begin{array}{*{20}c}
p_1'=-\frac{3(p_1+\rho_1) \left({\mathcal M}+4 \pi R^3
\left(p_1+\frac{\rho_2}{3}\right)\right)}{2  R \left(2\pi R-{3
{\mathcal M}}\right)},~~~~{\mathcal M}'=4\pi  R^2 (\rho_1+\rho_2),\\
\rho_2'=-\frac{6 \rho_2 \left({\mathcal M}+4 \pi R^3
\left(p_1+\frac{\rho_2}{3}\right)\right)}{ R \left(2\pi R-{3
{\mathcal M}}\right)},~~~~~~~~~
~~~~~~~~~~~~~~~~~~~~~~~~~~~~~~~~~~~
\end{array}} } \right. ~~~R_i<R<R_f,\\
\rho_2'=-\frac{6 \rho_2 \left({\mathcal M}+4 \pi R^3
\frac{\rho_2}{3}\right)}{ R \left(2\pi R-{3 {\mathcal
M}}\right)},~~~{\mathcal M}'=4\pi  R^2 (\rho_2+\rho_3),~~~~R>R_f.
\end{array}} } \right.
\label{bigsyst}
\ee
where the baryonic matter $\rho_3$ is considered as some external
matter density. According  to (\ref{bigsyst}), there are two
equations for the pressure and ``dark radiation'' density inside
the eicheon and a single equation for ``dark radiation'' density
outside the~eicheon.

As is shown in the upper panel of Figure~\ref{fig:fig3},
the~eicheon without galactic disk and bulge contributes at a small
distance, and~the dark radiation contributes at large distances.
The density of dark radiation depends on the eicheon structure,
which was considered in the previous section. It is convenient to
introduce  a universal quantity of a dark radiation density for
the Milky Way Galaxy at the radius of a photon sphere $R=3/2$,
which almost does not depend on eicheon structure, namely
$\rho_2^*\equiv\rho_2(3/2)=9.3\times 10^{-32}=2.1\times
10^{-25}$~kg/m$^3$. Moreover, it remains a single parameter
because the eicheon mass in the dimensionless units equals
$\mathbb{M}=2\pi/3$. Thus,  a~DM tail is reproduced by virtue of
the universal equations

\be
\rho_2'=-\frac{3 \rho_2 \left({\mathcal M}+4 \pi R^3
\frac{\rho_2}{3}\right)}{\pi R \left(R-\frac{3 {\mathcal M}}{2 \pi
}\right)},~~~{\mathcal M}'(R)=4\pi  R^2\rho_2,~~{\mathcal
M}(3/2)=2\pi/3,~ \rho_2(3/2)=\rho_2^*,~~R>3/2.
\ee

\begin{figure}[H]
\includegraphics[width=9.cm]{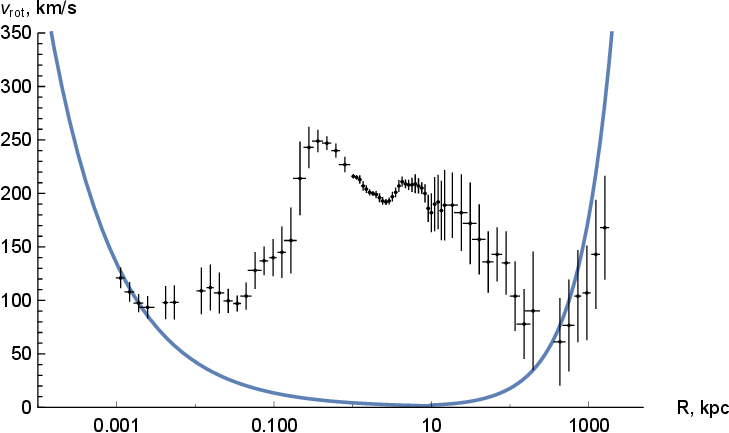}
\includegraphics[width=9.cm]{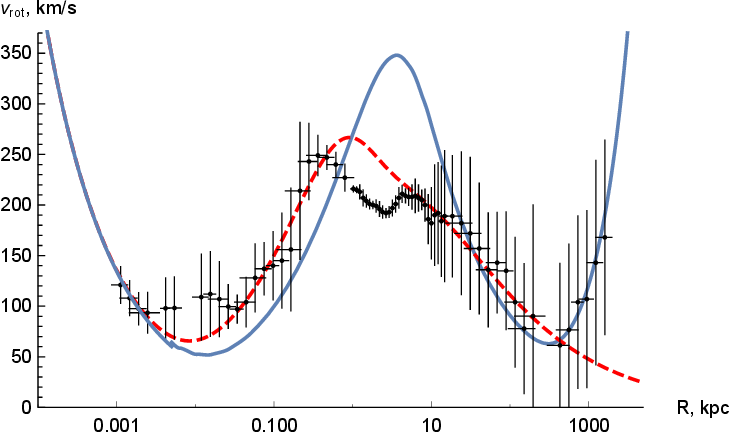}
\caption{(\textbf{Left panel}) The calculated rotational curve of
the Milky Way Galaxy from Ref.~\cite{polar2022}, which includes
contributions of the eicheon and dark radiation. (\textbf{{Right
panel}}) Rotational curve taking into account the baryonic matter
by (\ref{55})--(\ref{77}). Dashed line represents DM modelling  by
NFW profile, and~baryonic matter by (\ref{55})--(\ref{77}),
but~using another parameters. The~result of observations with the
error bars are taken from Ref.~\cite{Sofue}.}
    \label{fig:fig3}
\end{figure}

That is a spherically symmetric model where the amount of dark
radiation is adjusted to fit the observations. The~rotation
velocity is calculated according to~\cite{polar2022}
\be
v_{rot}=\sqrt{\frac{R}{2B}\frac{dB}{dR}}=
\sqrt{-\frac{R}{p_2+\rho_2}\frac{dp_2}{dR}}=\frac{1}{2}\sqrt{-\frac{R}{\rho_2}\frac{d\rho_2}{dR}},
\label{vrott}
\ee
where the last equality of (\ref{vrott})   says that the  dark
radiation $\rho_2$ serves a ``reference fluid'' because satisfies
continuity Equation~(\ref{eqB}) rewritten in the form of
\be
\frac{d\rho_2}{dR}+\frac{2}{B}\frac{dB}{dR}\,\rho_2=0.
\ee

To consider the baryonic matter, one could smear a baryonic
galactic disk on a sphere and view the resulting mass density as
some external non-dynamical density in the TOV equations for the
eicheon and dark radiation. This external density creates an
additional gravitational~potential.

Let us consider the surface density of matter in a galactic disk:
\be
\wp=\frac{M_D }{2\pi R_D^2 }e^{-R/R_D},
\label{55}
\ee
and write the mass $dM$ corresponding to the radial distance $dR$
\be
dM=\frac{M_D}{ R_D^2 }e^{-R/R_D}\,RdR= \frac{M_D}{
R_D^2R}e^{-R/R_D} \,R^2dR.
\label{66}
\ee
According to (\ref{66}), the~smeared 3-dimensional density has the
form:
\be
\rho_3=\frac{M_D}{4\pi R_D^2R}e^{-R/R_D}.
\label{77}
\ee

The result of the calculations for the Milky Way Galaxy rotational
curve is shown in the lower panel of Figure {\ref{fig:fig3}}.
As~one can see, the~simple model with smeared disk describes the
baryonic matter roughly, but~the observed rotational curve has a
more complicated structure.

\section{Andromeda~Galaxy}
\label{andr}

Andromeda Galaxy (M31) is nearest to the Milky Way Galaxy and is
situated at a distance $\sim$800~kpc. For~M31, there are no small
distances data  $\sim$0.01~kpc, allowing us to identify a compact
object in the center explicitly.  Indeed, the~situation is more
complicated because cluster B023--G078 of M31 hosts one more black
hole $\sim$$10^5~M_\odot$ \cite{1999ApJ,Bender2005}. For~the
correct description, we need to apply a solution with two
eicheons. This problem seems complicated, and~we leave it for the
future, considering only one central eicheon with the mass
$10^8~M_\odot$ \cite{1999ApJ,Bender2005}.
 Calculations are analogous to that of the previous~section.

The results of modeling are shown in Figure~\ref{fig:fig4A}.
The~dimensionless parameter equaled the dark radiation density in
the units of $M_p^2\,r_g^{-2}$ at a photon sphere radius is
$\rho_2(3/2)=6.2\times 10^{-28}=2.3\times 10^{-24}$~kg/m$^3$. This
value is greater than that for the Milky Way Galaxy, i.e.,~these
suggest the greater mass of central eicheon and the greater dark
radiation density at a radius of the eicheon photon sphere.
After~introducing the baryonic matter by smearing galactic disk
(\ref{55})  over a sphere
 we have the curve shown in a lower panel of Figure~\ref{fig:fig4A}.

 \begin{figure}[H]
\includegraphics[width=9.cm]{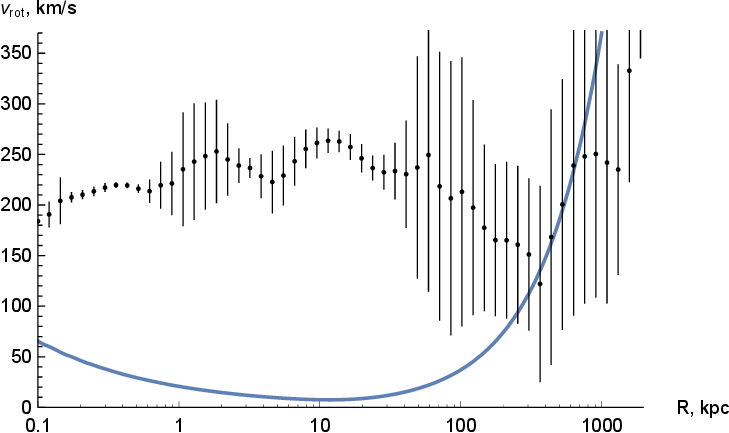}
\includegraphics[width=9.cm]{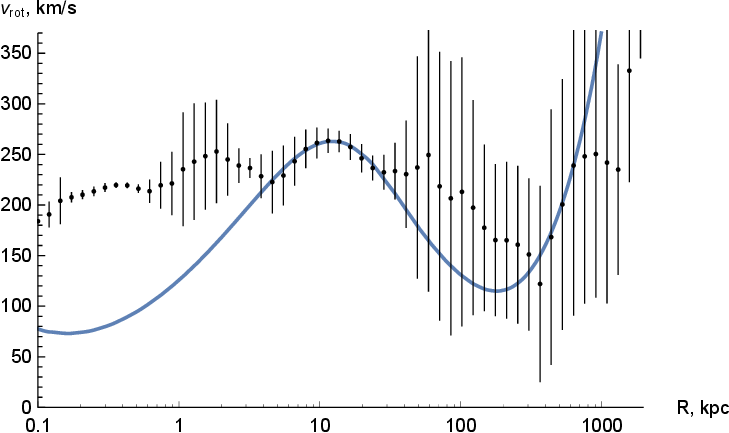}
\caption{(\textbf{Left panel}) Andromeda Galaxy rotational curve,
which includes contributions of the eicheon and dark radiation
only. (\textbf{{Right panel}}) Rotational curve taking into
account the baryonic matter by (\ref{55})--(\ref{77}). The~result
of observations with the deviations bars are taken from
Ref.~\cite{Sofue1}.}
\label{fig:fig4A}
\end{figure}

\section{Discussion and~Conclusions}

We have shown that the F-type vacuum polarization could explain
DM, which mimics a sort of ``dark radiation''. Namely the presence
of ECO, or~\textit{ {eicheon}
}, in~the center of the galaxy provides a nonsingular solution for
dark radiation. The~eicheon resembles a black hole for an external
observer but has no horizon. Our model is spherically symmetric.
However, the~appropriate approximation of the distribution of
baryonic matter in a galaxy by the disk smearing over a sphere
allows for obtaining the qualitative agreement of the rotational
curves with the observed~ones.

Still,  in~the spreading of a galactic disk, we overestimated a
baryonic matter. Usually, it is supposed that DM begins to play a
role from a few kpc, but~according to the above consideration, the
contribution of dark radiation becomes considerable  only at tens
kpc. For~example,  in~Figure~\ref{fig:fig3} (red dashed curve) we
represent DM modelling by Navarro Frenk White (NFW)
profile~\cite{Navarro1996}, where left and right edges of the
central fold of the plot are approached  by baryonic matter and
NFW profile~respectively.

For M31, we are not able to obtain an amount of DM needed in the
region of 10--100~kpc. We conjecture that, if~there is not only a
central eicheon in the galaxy but a number of eicheons, one could
glue dark radiation tails to every eicheons and create a needed
amount.

Let us remind the principles of calculation. We have considered
the vacuum polarization of F-type in CUM (\ref{int1}) and find
that it has a radiation-like equation of state. Then, we solve the
TOV equation for incompressible fluid and dark radiation and
obtain a nonsingular solution. To~consider the baryonic matter, we
smear a galactic disk and use the resulting density as some
external density. Interestingly, each galaxy's
 dark radiation tail can be described by a single
parameter: density of dark radiation at the radius of a photon
sphere of the eicheon. The~numerical value of this density for the
Milky Way Galaxy is $2.1\times 10^{-25}$~kg/m$^3$, and~ for
Andromeda Galaxy it is $2.3\times 10^{-24}$~kg/m$^3$.

These values could be compared with the spatially uniform residual
energy density of vacuum fluctuations, which remain after
compensation of its main part by the constant in the Friedman
equation~\cite{Haridasu}. It is of the order of critical density
$\sim$$\times 10^{-26}$~kg/m$^3$. Certainly, we use the amount of
the dark radiation at a photon radius of the eicheon $R=3/2r_g$.
Still, this amount rapidly decreases at $R>3/2 r_g$ and increases
at $R<3/2 r_g$. In~this light, it is interesting to obtain a
general picture of matter structure formation in the universe by
the solution of the system of the equations for the perturbations
of the metric and the matter, including  vacuum polarisation of
both types~\cite{polar2022}.

\appendix

\section{TOV Equation for a Mixture of Ordinary and Dark Fluids}
\label{dertov}
Each of the fluids obeys the equation of the hydro-static
equilibrium~\cite{wein}:
\be
\frac{B^\prime}{B}=-\frac{2 p_1^\prime}{\rho_1+p_1},
\label{fr0}
\ee
whereas the equations for gravitational field give~\cite{wein}
\bea
\left(\frac{R}{A}\right)^\prime=1-8\pi G\rho R^2,\label{fr1}\\
-1+\frac{R}{2A}\left(-\frac{A^\prime}{A}+\frac{B^\prime}{B}\right)+\frac{1}{A}=-4\pi
G(\rho-p)R^2,
\label{fr2}
\eea
where $p=p_1+p_2$, and~$\rho=\rho_1+\rho_2$. Solution of the
Equation~(\ref{fr1}) is written formally as
\be\
A=\frac{1}{1-2G {\mathcal M}/R}\,,
\label{aa1}
\ee
where $\mathcal M(R)$ is given by (\ref{mmr}) Expressing
$B^\prime/B$ from (\ref{fr0}), $A$, $A^\prime$ from (\ref{aa1}),
(\ref{fr2}) and substituting them into (\ref{fr1}) gives
\be
p_1^\prime=-(p_1+\rho_1)\frac{G \left(\mathcal M+4 \pi  R^3
p\right) }{R (R-2 G \mathcal M)}.
\ee

 {The} analogous equation holds for the second fluid.

\end{document}